\documentclass{emulateapj}
\usepackage{amssymb}
\usepackage{amsmath}
\usepackage{epstopdf}
\usepackage{natbib}
\usepackage[colorlinks, citecolor=blue]{hyperref}
\usepackage{hyperref}
\usepackage[all]{hypcap}

\begin{document}

\title{A Triple-Energy-Source Model for Superluminous Supernova iPTF13ehe}
\author{S. Q. Wang\altaffilmark{1,2}, L. D. Liu\altaffilmark{1,2}, Z. G. Dai\altaffilmark{1,2}, L. J. Wang\altaffilmark{3}, and X. F. Wu\altaffilmark{4,5,6}}
\affil{\altaffilmark{1}School of Astronomy and Space Science, Nanjing University, Nanjing 210093, China; dzg@nju.edu.cn}
\affil{\altaffilmark{2}Key Laboratory of Modern Astronomy and Astrophysics (Nanjing University), Ministry of Education, China}
\affil{\altaffilmark{3}Key Laboratory of Space Astronomy and Technology, National Astronomical Observatories, Chinese Academy of Sciences, Beijing 100012, China}
\affil{\altaffilmark{4}Purple Mountain Observatory, Chinese Academy of Sciences, Nanjing, 210008, China}
\affil{\altaffilmark{5}Chinese Center for Antarctic Astronomy, Chinese Academy of Sciences, Nanjing, 210008, China}
\affil{\altaffilmark{6}Joint Center for Particle Nuclear Physics and Cosmology of Purple Mountain
Observatory-Nanjing University, Chinese Academy of Sciences, Nanjing 210008, China}

\begin{abstract}

Almost all superluminous supernovae (SLSNe) whose peak magnitudes are $\lesssim -21$ mag
can be explained by the $^{56}$Ni-powered model, magnetar-powered (highly magnetized pulsar)
model or ejecta-circumstellar medium (CSM) interaction model.
Recently, iPTF13ehe challenges these energy-source models, because the
spectral analysis shows that $\sim 2.5M_\odot$ of $^{56}$Ni have been synthesized but
are inadequate to power the peak bolometric emission of iPTF13ehe, while the rebrightening
of the late-time light-curve (LC) and the H$\alpha$ emission lines indicate that the ejecta-CSM
interaction must play a key role in powering the late-time LC. Here we propose a triple-energy-source model,
in which a magnetar together with some amount ($\lesssim 2.5M_\odot$) of $^{56}$Ni may power the early LC of iPTF13ehe
while the late-time rebrightening can be quantitatively explained by an ejecta-CSM
interaction. Furthermore, we suggest that iPTF13ehe is a genuine core-collapse supernova
rather than a pulsational pair-instability supernova candidate.
Further studies on similar SLSNe in the future would eventually shed light on
their explosion and energy-source mechanisms.

\end{abstract}


\section{Introduction}
\label{sec:Intro}

Supernovae (SNe) are exceptionally brilliant stellar explosions in the Universe.
Over the last decade, dozens of superluminous SNe (SLSNe)
 whose peak magnitudes are $\lesssim -21$ mag \citep{Gal2012}
 have attracted intense interest from both observational and theoretical astronomers.
SLSNe are the rarest class of supernovae discovered so far
 and have been classified into types I (hydrogen-poor) and II (hydrogen-rich).
 Comparing the results of \citet{Cooke2012}, \citet{Qui2013}, and \citet{Taylor2014},
the ratio of the explosion rates of SLSNe to core-collapse SNe (CCSNe)
 is $\sim(1-4)\times10^{-3}$.

Discriminating among the energy-sources powering the SLSNe is rather tricky.
 However, these energy sources should leave their imprints on the light curves (LCs) and the spectra of these SLSNe.
Only very few SLSNe (e.g., SN 2007bi) can be regarded as
 the ``pair instability SNe (PISNe)'' \citep{Bar1967,Rak1967,Heg2002,Heg2003}
 and explained by the $^{56}$Ni-powered model \citep{Gal2009}\footnote{However, \citet{Des2012}
 and \citet{Nich2013} disfavored the $^{56}$Ni-powered model in explaining SN 2007bi and
 other similar events.},
while the majority of SLSNe cannot be solely powered by $^{56}$Ni \citep{Qui2011,Inse2013} but
 can be explained by the ejecta-circumstellar medium (CSM) interaction model \citep{Che2011,Gin2012} or
 magnetar-powered (ultra-highly magnetized pulsar) model \citep{Kas2010,Woos2010,Des2012,Inse2013,Wang2015a}.
 Because of the absence of emission lines indicative of the ejecta-CSM interaction,
the magnetar model is preferred in explaining almost all
 SLSNe-I \citep[e.g.][]{Inse2013,Nich2013,How2013,McC2014,Vre2014,Nich2014,Wang2015a,Met2015,Dai2016,Kas2016,Wang2016}.
Meanwhile, the ejecta-CSM interaction model works well in explaining SLSNe-II,
 especially superluminous SNe IIn \citep{Smi2007,Mor2013a}.

In principle, $^{56}$Ni should be taken into account in light-curve modeling for all SLSNe.
From the photometric aspect, however,
 SLSNe's $^{56}$Ni yields are usually rather small \citep[e.g.][]{Inse2013} and the luminosities provided by $^{56}$Ni are
 outshone by magnetars or ejecta-CSM interaction, except for some PISN candidates that might be mainly powered by $^{56}$Ni.
Besides, $^{56}$Ni yields are difficult to
 be determined precisely owing to the lack of late-time spectra for most SNe.
 Hence, excluding a few PISN candidates, previous
 modelings for most LCs of SLSNe-I omitted the contributions of $^{56}$Ni.

There are some more complicated cases: for some SLSNe-II, e.g. SN~2006gy,
 adding several $M_\odot$ of $^{56}$Ni can make the fitting more reliable,
the ``$^{56}$Ni+interaction" model has been employed \citep{Agn2009,Cha2012,Cha2013}.
No SLSN has involved with all three energy sources so far;
\citet{Wang2015b} suggested that some luminous SNe Ic with
peak magnitudes $\sim -20$ mag must be explained by a unified scenario containing the
 contributions from $^{56}$Ni and a magnetar (``$^{56}$Ni+interaction" model) since they cannot be
  powered solely by $^{56}$Ni \citep[see also][]{Gre2015} but
the contribution from $^{56}$Ni cannot be neglected \citep[see also][]{Ber2016}.

Recently, an SLSN-I at redshift $z=0.3434$, iPTF13ehe, might change the conclusions above.
\citet{Yan2015} analyzed the light curve and spectra of iPTF13ehe and found that
 the peak bolometric luminosity of iPTF13ehe is $\sim 1.3\times10^{44}$ erg s$^{-1}$.
 If this SLSN was powered by $^{56}$Ni, more than $(13-16) M_\odot$ of $^{56}$Ni must be synthesized
 to power the peak bolometric emission \citep{Yan2015}.
However, the late-time spectral analysis indicated that the mass of $^{56}$Ni
 synthesized is $\sim$ 2.5 $M_\odot$.
This amount of $^{56}$Ni, although it cannot be ignored, is inadequate to power the peak bolometric luminosity.
 Therefore, \citet{Yan2015} argued that the early LC must be powered by multiple power sources.

Furthermore, the presence of H$\alpha$ emission lines in the late-time spectrum and the rebrightening
 in the late-time LC show an unambiguous signature for interactions between
the SN ejecta and a hydrogen-rich CSM.
Hence, any model with single power source or double power sources is challenged in explaining the LC of iPTF13ehe.
A triple-power-source model should be seriously considered.

In this paper, we propose a triple-energy-source model for iPTF13ehe and try to understand its explosion mechanism and LC.
This paper is organized as follows.
In Sections \ref{sec:early} and \ref{sec:late-time}, we fit the early-time and late-time LC using double and
triple energy-source models, respectively.
 In Sections \ref{sec:dis} and \ref{sec:con}, we discuss and conclude our results.

\section{Modeling the Early-Time LC}
\label{sec:early}

We start by studying the early LC
 and determining the power sources of iPTF13ehe.
 As pointed out by \citet{Yan2015}, $\sim 2.5 M_\odot$ of $^{56}$Ni were synthesized while
$\gtrsim 13-16 M_\odot$ of $^{56}$Ni are required to account for the LC peak.
Therefore, there should be other energy sources to aid the high peak-luminosity of iPTF13ehe.
The early-time spectra shows no H emission lines, indicating that the ejecta-CSM
 interaction can be neglected in the early-time luminosity evolution. Hence,
 the energy source aiding the SN might be a magnetar spin-down input.
 The possibility that iPTF13ehe is a CCSN and can be aided by a magnetar was also pointed out by \citet{Yan2015},
  although they didn't provide further analysis.
 Considering the contribution from $^{56}$Ni that cannot be neglected, the most reasonable model
 should contain contributions from both $^{56}$Ni and a magnetar.
\citet{Wang2015b} have constructed such a unified model for SNe Ic
  and explained three luminous SNe Ic. We here employ this model to
 fit the LC of iPTF13ehe.

Supposing that the ratio of the $r$-band flux to bolometric flux remains nearly constant around the peak and in the late-time,
 we scale the $r$-band luminosities to obtain the bolometric luminosities. The scaled
bolometric luminosities are rough, but they can be regarded as an approximation of real values.
These data are plotted in Fig. \ref{fig:fit}. In this section, we use some models to fit the early-time LC.
The photospheric velocity and the optical opacity $\kappa$ are
 fixed to $\sim13,000$ km s$^{-1}$ \citep{Yan2015} and $0.2$ cm$^{-2}$ g$^{-1}$, respectively.

\subsection{$^{56}$Ni Model}

\begin{figure}[htbp]
\begin{center}
   \includegraphics[width=0.53\textwidth,angle=0]{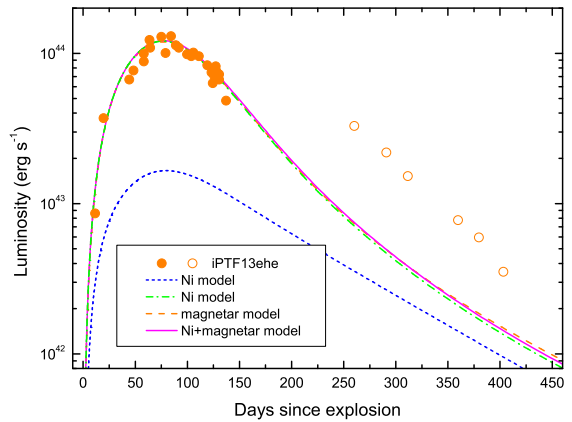}
   \caption{Modeling iPTF13ehe's early-time LC using $^{56}$Ni, magnetar, and hybrid ($^{56}$Ni + magnetar) models.
   Data are obtained from \citet{Yan2015}. The early-time and late-time bolometric luminosities scaled from $r$-band ones
 are shown as the orange filled circles and orange unfilled circles, respectively.
The dash-dotted line is the LC powered by 19 $M_\odot$ of $^{56}$Ni;
the dashed line is the LC powered by a magnetar;
the solid line is the LC powered by $^{56}$Ni and a magnetar.
The short dashed line is the LC powered by 2.5 $M_\odot$ of $^{56}$Ni with full trapping of gamma-rays.
The other model parameter values are shown in the text.}
   \label{fig:fit}
\end{center}
\end{figure}

The first of the models employed here is the $^{56}$Ni-powered model.
 We find that when the ejecta mass $M_{\rm ej}=35~M_\odot$,
 the gamma-ray opacity $\kappa_{\gamma}\simeq0.018$ cm$^{-2}$ g$^{-1}$,
 and the $^{56}$Ni mass $M_{\rm Ni} \sim19~M_\odot$,
 the theoretical LC can matches the observational data at early times ($t\lesssim137$ days), as can be seen in Fig. \ref{fig:fit}
 (the dash-dotted line). The $^{56}$Ni mass obtained here is slightly larger than the value $\sim (13-16)M_\odot$ inferred by \citet{Yan2015}.

However, the ratio of $M_{\rm Ni}$ ($\sim19~M_\odot$) to $M_{\rm ej}$ (35 $M_\odot$)
 is $\sim 50\%$, significantly exceeding the upper limit of the values for
  CCSNe (the ratio of $^{56}$Ni mass to the ejecta mass is $\lesssim 20\%$ \citep{Ume2008}).
 Therefore, the $^{56}$Ni-powered model can be excluded in explaining the data.

\subsection{Magnetar Model and Magnetar+$^{56}$Ni Model}

In the magnetar-powered model, if the
 the initial rotational period of the magnetar $P_{0}=2.45$~ms,
 the surface magnetic field $B=8\times10^{13}$ G,
 $\kappa_{\gamma}\simeq0.018$ cm$^{-2}$ g$^{-1}$, then the dashed line in Fig. \ref{fig:fit}
is plotted to obtain an equivalently good fitting.
Since the spectral analysis has inferred that the SN explosion produced $\sim2.5~M_\odot$ of $^{56}$Ni,
 we take into account the contribution from $^{56}$Ni and use the hybrid model, in which
 the initial period of the magnetar $P_{0}$ is larger slightly, $\sim2.55$~ms.
From the fit, we can get other two important physical quantities relevant to the next section, i.e., the rise time
 $t_{\rm r}\sim81$ days and the kinetic energy
 of the SN explosion, $E_{\rm SN}=(3/10)M_{\rm ej}v^2\sim3.5\times10^{52}$ erg.

\subsection{Further Analysis for the Parameters of iPTF13ehe}

A caveat on the degeneracy of model parameters should be clarified here.
The effective light-curve timescale $\tau_{m}=(2\kappa M_{\rm ej}/\xi vc)^{1/2}$ \citep{Arn1982} determines the
 width and therefore the rise time of the LCs of SNe,
where $c$ is the speed of light and $\xi=4\pi^3/9\simeq13.8$ is a constant.
If the value of $\kappa M_{\rm ej}/v$ is invariant,
 LCs around the peak are the same. Although the photospheric velocity $v$ of iPTF13ehe is fixed to 13,000 km s$^{-1}$,
 the degeneracy of $\kappa$ and $M_{\rm ej}$ is still difficult to be broken, because larger $\kappa$ requires smaller
 $M_{\rm ej}$ for the same level of fittings.
 The optical opacity of H- and He-deficient ejecta has been
 assumed to be 0.06 cm$^2$~g$^{-1}$ \citep[e.g.,][]{Val2011,Lym2016}, 0.07 cm$^2$~g$^{-1}$ \citep[e.g.,][]{Tad2015,Wang2015b},
 0.08 cm$^2$~g$^{-1}$ \citep[e.g.,][]{Arn1982,Maz2000}, 0.10 cm$^2$~g$^{-1}$ \citep[e.g.,][]{Nug2011,Inse2013,Whe2015}, and 0.2 cm$^2$~g$^{-1}$ \citep[e.g.,][]{Kas2010,Nich2014}.
We assume that $\kappa=$0.2 cm$^2$~g$^{-1}$ throughout this paper and get the ejecta mass $M_{\rm ej}\sim35M_\odot$.
If we assume a smaller $\kappa$, e.g., $\kappa=$0.1 cm$^2$~g$^{-1}$, then the ejecta mass becomes larger, $M_{\rm ej}\sim70M_\odot$. The
 latter is consistent with the lower limit derived by \citet{Yan2015}.
 Provided that iPTF13ehe is a genuine core-collapse SN,
the ejecta with mass of $35M_\odot$ is preferred.



\section{Modeling the Late-Time LC}
\label{sec:late-time}

We have demonstrated that the hybrid model containing the contributions from a magnetar and $^{56}$Ni
 can well fit the early-time LC of iPTF13ehe,
but the late-time rebrightening cannot be explained by such a hybrid model.
\citet{Yan2015} found that the late-time spectrum of the SN displayed H$\alpha$ emission lines which are usually present in SNe IIn
 and inferred that a Hydrogen shell ejected about 40 years before the SN explosion, locating
 at a distance of $\sim4\times10^{16}$ cm from the SN progenitor. When the SN ejecta
 run into the shell, an interaction was triggered and thus the $r$-band LC was rebrightened
 and the H$\alpha$ emission was prompted.
Therefore, both the photometric and spectral features
 indicate that the interaction between the ejecta and CSM must play an important role in
 powering the late-time LC \citep{Yan2015}.

However, the exact time of the emergence of the ejecta-CSM interaction is uncertain due to the lack of
 the photometric data at $\sim60-170$ days and the spectral data at $13-251$ days after the peak.
Nevertheless, the fact that there is no rebrightening in $\sim 56$ days after the peak provides
 a lower limit on the interaction onset time, while the rebrightening at $\sim 180$ days after the peak offers an upper limit.
Therefore, we suppose that the rebrightening started at $\sim56-180$ days after the peak.
Adopting $t_{\rm r}\sim81$ days, we conclude that the interaction should
 be triggered between $\sim137-261$ days after the SN explosion. Hence,
 it is reasonable to suppose that the onset time of interaction between the ejecta and the CSM-shell $t_{\rm i}$ is $\sim140$~days.

\begin{figure}[htbp]
\begin{center}
   \includegraphics[width=0.53\textwidth,angle=0]{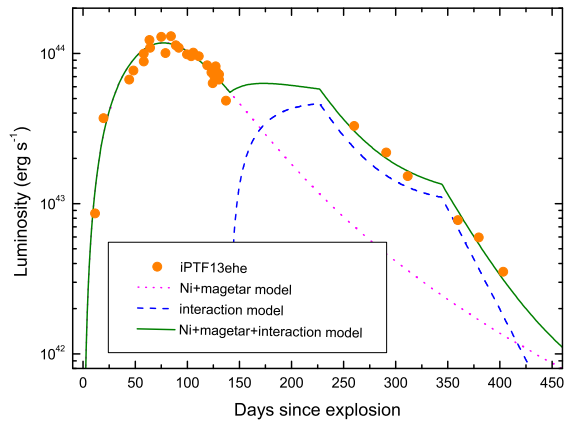}
   \caption{Modeling iPTF13ehe's whole LC by using the triple-energy-source ($^{56}$Ni + magnetar + interaction) model.
   Data are obtained from \citet{Yan2015}. The bolometric luminosities scaled from $r$-band ones
 are shown as the orange filled circles. The dotted line is the LC powered by $^{56}$Ni and a magnetar;
the dashed line is the LC powered by the ejecta-CSM interaction;
the solid line is the LC powered by the triple-energy-source (``$^{56}$Ni+magnetar+ejecta-CSM interaction") model.
The values of other parameters are shown in the text.}
   \label{fig:fit2}
\end{center}
\end{figure}

According to Equations (14)-(16) and (20) of \citet{Cha2012}, we
reproduce the late-time LC powered by the ejecta-CSM interaction, seen in Fig. \ref{fig:fit2}.
The fit parameters are the optical opacity of the H-shell $\kappa'$=0.34~cm$^2$ g$^{-1}$,
the ejecta mass $M_{\rm ej}=35 M_\odot$, the SN kinetic energy $E_{\rm SN} = 3.5 \times 10^{52}$ erg (these two parameters are derived
from the previous section),
the slope of the inner density profile of the ejecta $\delta=0$, the slope of the outer density profile of the ejecta $n=7$,
the power-law exponent for the CSM density profile $s=0$ (corresponding to a uniform-density shell),
the density of the CSM $\rho_{\rm CSM}=10^{-14}$ g cm$^{-3}$, the mass of the CSM $M_{\rm CSM}=27M_\odot$,
 the onset time of interaction between the ejecta and the CSM-shell $t_{\rm i}\sim140$~days,
the conversion efficiency from kinetic energy to radiation $\epsilon=0.05$ that is comparable to
 the typical value $\epsilon\sim0.1$ \citep{Mor2013b}.

The density of the CSM shell in our model is one order of magnitude larger than the value derived by \citet{Yan2015}.
Since the shock-ionized CSM shell is probably optically thin to the Thomson
scattering \citep{Yan2015}, $\tau_{\rm Thomson}=\kappa'\rho{w}\leq1$, where $w$ is the
width of the shell, $\kappa'(M_{\rm CSM}/4{\pi}R^2{w}){w}=\kappa'(M_{\rm CSM}/4{\pi}R^2)\leq1$,
so $M_{\rm CSM}\leq4{\pi}R^2/\kappa'$. \citet{Yan2015} supposed that the interaction time is 251 days after the peak, and derived that $R\simeq4\times10^{16}$~cm.
 If $w=0.1R$, and $\kappa'=0.34$ cm$^2$~g$^{-1}$, then $M_{\rm CSM}\leq30M_\odot$, $n=\rho/m_{\rm H}\leq1/\kappa'{w}m_{\rm H}\sim4\times10^8$ cm$^{-3}$ \citep{Yan2015}, and $\rho\leq0.74\times10^{-15}$ g cm$^{-3}$.

This difference is partly caused by the uncertainty of the onset time of the ejecta-CSM interaction.
We suppose that the interaction time is $\sim$60 days after the peak and then have $R=vt\sim1.6\times10^{16}$~cm;
 if $w=0.1R$, and $\kappa'$=$0.34$ cm$^2$~g$^{-1}$, then $M_{\rm CSM}\leq5M_\odot$, $n\leq1.1\times10^9$ cm$^{-3}$, and
$\rho\leq1.84\times10^{-15}$ g cm$^{-3}$. As pointed out by \citet{Yan2015}, the possibility that the CSM shell might be partly ionized rather than fully ionized  cannot be excluded.
If the ratio of ionized CSM to total CSM is $\sim 1/5.4$,
 then $\rho_{\rm CSM}\sim10^{-14}$ g cm$^{-3}$, and $M_{\rm CSM}\sim27M_\odot$, being
 consistent with the values of our fitting parameters.

As shown in Fig. \ref{fig:fit2}, the LC powered by the
 triple power-source model (``$^{56}$Ni+magnetar+ejecta-CSM interaction" model)
 is well consistent with the data. If we ignore the contribution from $^{56}$Ni,
 the LC can be reproduced by the magnetar and ejecta-CSM interaction.

We caution that the bolometric LC is obtained by scaling the $r$-band LC and therefore is not
 accurate enough. The uncertainties in the bolometric luninosities prevent us from getting precise estimate of the model parameters.
 Nevertheless, this does not change the facts that the early-time LC is mainly powered by a magnetar rather than $^{56}$Ni and
  that the late-time LC requires adding the ejacta-CSM interaction. More accurate bolometric corrections can pose more
 stringent constraints on the parameters.

\section{Discussions}
\label{sec:dis}

The $^{56}$Ni production is usually ineffective after the explosion of a neutrino-driven CCSN.
Radioactive $^{56}$Ni in a neutrino-driven CCSN is mainly synthesized in two ways:
a neutrino-driven wind and an explosive nuclear burning of silicon (and other intermediate-mass elements)
 heated to a temperature exceeding $5 \times 10^9$ K.
 The neutrino-driven wind
 can produce $\sim 10^{-3} M_\odot$ of $^{56}$Ni and can be neglected;
 the amount of $^{56}$Ni synthesized in shock-heating silicon shell
 is related with the shock energy and therefore the kinetic energy of a CCSN. When the
 kinetic energy is $\sim 10^{51}$ erg, the mass of $^{56}$Ni is
 $\sim (0.05-0.1) M_\odot$.
If the neutron star left behind the explosion is
 a millisecond magnetar, it can help the ejecta
  synthesize additional $^{56}$Ni up to $\sim 0.2 M_\odot$ \citep{Suwa2015}.
  The total amount of $^{56}$Ni from these processes
 is $\lesssim 0.3 M_\odot$, significantly less than 2.5 $M_\odot$.

Therefore, it seems that the model containing the contributions from 2.5 $M_\odot$ of $^{56}$Ni and a magnetar
 is problematic. However, the possibility that the $^{56}$Ni mass of iPTF13ehe
 has been overestimated cannot be completely excluded.
 We can get some indirect clues from SN 2007bi: \citet{Gal2009} proposed that
 the amount of $^{56}$Ni synthesized by SN 2007bi is $\sim 5 M_\odot$ by matching the low-quality
nebular-phase spectrum and concluded that SN 2007bi is a PISN,
 while \citet{Des2012} showed that the PISN model must produce cool photospheres,
red spectra and narrow-line profiles which conflict with observational properties
 of SN 2007bi, and concluded that SN 2007bi might be powered by a magnetar.
 This example indicates that the simple spectral analysis is not enough to
  determine the precise value of $^{56}$Ni yield and
a more detailed analysis is needed.
In this paper, we can regard $\sim 2.5 M_\odot$ as the upper limit
 of $^{56}$Ni yield of iPTF13ehe.

We note that SN ejecta-CSM interaction might also be a promising
 scenario for explaining the early-time LC of iPTF13ehe.
If the progenitor experiences some eruptions and expels some discrete shells before its explosion,
 the ejecta can collide these shells at different epoches and the interactions between the ejecta and these shells
  can power the early-time LC.
 The intermittent spectral observations at early times might miss the spectral features.
 In this scenario, the putative millisecond magnetar can be removed and the central remnant is therefore a black hole.
However, how does the explosion synthesize $\sim 2.5M_\odot$ of $^{56}$Ni in this scenario
remains an open problem.

\section{Conclusions}
\label{sec:con}

We have analyzed and fitted the LC for iPTF13ehe which is an SLSN I with
 peak bolometric luminosity $\sim1.3\times10^{44}$ erg s$^{-1}$.
Using the $^{56}$Ni-powered model, we found that $\sim 19 M_\odot$
 of $^{56}$Ni are required, confirming the conclusion reached by \citet{Yan2015}.
Spectral analysis performed by \citet{Yan2015} shows that the SLSN synthesized $\sim 2.5 M_\odot$ of
 $^{56}$Ni but this amount of $^{56}$Ni is inadequate to power the peak luminosity.

We proposed that the most reasonable model accounting for the early-time LC
 is the magnetar-dominated model containing the contributions
 from a magnetar with $P_0 = 2.45-2.55$~ms and $B = 8\times10^{13}$ G
 and $\lesssim 2.5M_\odot$ of $^{56}$Ni. The LCs reproduced by this model
are in good agreement with early-time observations of iPTF13ehe.
The energy released by the magnetar is dominant in powering the
 optical LC of iPTF13ehe.

While proposing this hybrid model containing triple energy-sources,
  we also note that this model is not unique model of explaining iPTF13ehe since
 the early LC might also be powered by the ejecta-CSM
 interaction and $^{56}$Ni.
Since the CCSNe usually produce $\lesssim 0.3 M_\odot$ of $^{56}$Ni,
 $\sim 2.5 M_\odot$ of $^{56}$Ni is problematic in both ``magnetar+$^{56}$Ni" model and
 ``interaction+$^{56}$Ni''. We regard $\sim 2.5 M_\odot$ as the upper limit and
 call for more detailed analysis.

An important spectral feature owned by iPTF13ehe is H$\alpha$ emission lines in its late-time spectrum.
These lines, together with the rebrightening of the late-time LC \citep{Yan2015},
 indicate that the interaction between
 the SN ejecta and the CSM has been triggered. The hybrid model involving $^{56}$Ni and magnetar cannot account
 for the late-time rebrightening. We find that the ejecta-CSM interaction model can
 explain the rebrightening and is the natural requirement accounting for the H$\alpha$ emission lines. Therefore,
 we use the model containing a magnetar, $^{56}$Ni, and ejecta-CSM interaction to fit the whole LC.

Finally, we discuss the explosion nature of iPTF13ehe.
The physical properties of the progenitors, the explosion mechanisms and energy-source mechanisms
 of SLSNe are still ambiguous and controversial \citep[see][for further analysis]{Des2012,Nich2013}.
As emphasized by \citet{Yan2015}, the massive Hydrogen-rich shell must have been expelled by the
 ``pulsational pair-instability (PPI)" mechanism \citep{Heg2003,Woos2007,Pas2008,Chu2009,Cha+Whe2012}.
The ejecta colliding with CSM shell contains $\lesssim2.5M_\odot$ of $^{56}$Ni and $\gtrsim3.5\times10^{52}$ erg
 of kinetic energy. This ejecta is not a shell following the hydrogen-rich shell, because
\citet{Heg2003} pointed out that the shell expelled by PPI contains no $^{56}$Ni and its kinetic energy
 is up to several times $10^{51}$ erg. The kinetic energy of the ejecta of iPTF13ehe is at least one order of magnitude
 higher than that of each PPI pulsed shell.
Therefore, we propose that iPTF13ehe itself is a genuine CCSN rather
 than a PPISN candidate, while the hydrogen-rich massive shell could be caused by a PPI pulse $\sim$
 20 years before the CCSN explosion ($\delta{t} \sim {v}t_{i}/v_{\rm CSM} \sim 13,000\;{\rm km/s}\times140\;{\rm days}/300\;{\rm km/s}$ $\sim$ 17 years, the velocity of the CSM shell $v_{\rm CSM}$ was inferred by \cite{Yan2015}).

\citet{Yan2015} estimated that at least $15\%$ of SLSNe-I may own such interaction features for late-time spectra.
We expect that future studies focusing on similar SLSNe should provide
 further insight into their explosion and energy-source mechanisms.

\acknowledgments
We thank an anonymous referee for valuable comments and constructive suggestions
that have improved our manuscript. This work was supported by the National Basic Research
Program (``973'' Program) of China (grant No.
2014CB845800) and the National Natural Science Foundation
of China (grant Nos. 11573014 and 11322328). X.F.W. is partially
supported by the Youth
Innovation Promotion Association (2011231), and the Strategic Priority
Research Program ``The Emergence of Cosmological Structures''
(grant No. XDB09000000) of of the Chinese Academy
of Sciences.

\end{document}